 \documentstyle[nato,epsfig]{crckapb} 

\newcommand{\ba}{\begin{eqnarray}}
\newcommand{\ea}{\end{eqnarray}}
\newcommand{\bmath}{\begin{mathletters}}
\newcommand{\emath}{\end{mathletters}}
\newcommand{\ban}{\begin{eqnarray*}}
\newcommand{\ean}{\end{eqnarray*}}
\newcommand{\tl}{\tilde{\ell}}

\begin{opening}
\title{RELATIVISTIC PSEUDOSPIN SYMMETRY and\protect\\
 the STRUCTURE of NUCLEAR STATES}

\author{A. Leviatan$^1$ and J.N. Ginocchio$^2$}
\institute{$^{1}$Racah Institute of Physics, The Hebrew University,\\
Jerusalem 91904, Israel\protect\\
$^{2}$Theoretical Division, Los Alamos National Laboratory,\\ 
Los Alamos, New Mexico 87545, USA}
\end{opening}

\runningtitle{RELATIVISTIC PSEUDOSPIN SYMMETRY and NUCLEAR STATES}

\begin{document}


\section{Introduction}

The concept of pseudospin symmetry \cite{aa,kth} 
is based on the empirical observation 
of quasi-degenerate pairs of certain normal-parity
shell-model orbitals with non-relativistic quantum numbers
\ba
\left (\,n_{r},\ \ell,\ j = \ell + {1\over 2}\,\right )\;\;\;
{\rm and}\;\;\;
\left (\,n_{r}-1,\ \ell + 2,\ j = \ell + {3\over 2}\,\right ) ~.
\label{psdoub}
\ea
Here $n_r$, $\ell$, and $j$ are the
single-nucleon radial, orbital, and total angular momentum quantum
numbers, respectively. The doublet structure, 
is expressed in terms of a ``pseudo'' orbital angular momentum
$\tilde{\ell}$ = $\ell$ + 1 and ``pseudo'' spin, $\tilde s$ = 1/2, 
coupled to $j = \tilde{\ell}\pm \tilde s$.
For example, $(2 p_{3/2},1f_{5/2})$ will have $\tilde{\ell}= 2$, etc. 
This pseudospin symmetry plays a central role in nuclei \cite{gl1} and 
only recently has it been shown to originate from a relativistic symmetry 
of the Dirac Hamiltonian \cite{gino,ami}. 
The members of the pseudospin doublet exhibit the following features. 
(i)~They have different angular momentum quantum numbers 
$(\ell,\,j)\leftrightarrow(\ell+2,\;j+1)$ for states with 
aligned/unaligned spin. 
(ii)~They have different radial wave functions and, in particular, 
different number of nodes $n_r\leftrightarrow n_r-1$. 
(iii)~They involve only normal-parity shell-model orbitals. 
The ``intruder'' levels with aligned spin and no nodes, {\it e.g.} 
$0g_{9/2},\;0h_{11/2},\;0i_{13/2}$ do not form quasi-degenerate 
doublets. In the present contribution we show that a natural explanation 
for all these features can be obtained by combining the relativistic 
attributes of pseudospin symmetry with known properties of Dirac bound 
states \cite{levgino2}.

\section{Dirac Wave Functions in Central Fields}

The eigenstates of a spherically symmetric Dirac Hamiltonian have the 
form $\Psi_{\kappa,m} = \left ( 
{G_{\kappa}\over r}\left [Y_{\ell}\,\chi\right]^{(j)}_{m}\,,\,
i{F_{\kappa}\over r}\left [Y_{\ell'}\,\chi\right]^{(j)}_{m}\,\right )$ 
where $G_{\kappa}(r)$ and $F_{\kappa}(r)$ 
are the radial wave functions of the upper and lower components 
respectively. The labels $\kappa=\mp(j+1/2)$ and $\ell' = \ell\pm 1$ 
for $j=\ell\pm 1/2$. 
As bound state solutions, both $G(r)$ and $F(r)$ vanish at $r=0$ and 
$r=\infty$ and the binding energy is positive $(M-E)>0$.
Further properties of these wave functions can be inferred from 
the following equation for the product $GF$ 
\ba
\left ( GF \right )' &=& A(r)\, F^2 - B(r)\, G^2 ~,
\label{GFprime}
\ea
where
$A(r) = \left [E + M + V_S(r) - V_V(r)\right ]$, 
$B(r) = \left [E - M - V_S(r) - V_V(r) \right ]$.
For relativistic mean fields relevant to nuclei, 
the scalar potential, $V_S(r)$, is attractive and the vector potential, 
$V_V(r)$, is repulsive. For typical values 
$A(r) > 0$, while $B(r)$ changes monotonically from $B(0)>0$ to 
$B(\infty)=(E-M)<0$. Since $GF$ vanishes at both end points.
we find by Eq.~(\ref{GFprime}) that it is 
an increasing negative function at large $r$, while at small $r$, 
$GF$ is a decreasing negative function for $\kappa<0$ and 
an increasing positive function for $\kappa>0$.
Futhermore, since $B(r)>0$ at the nodes of $F$ and $G$
(nodes can occur only in the region where the kinetic energy is positive), 
it follows that $GF$ is a decreasing function at the nodes of $F$, and an 
increasing function at the nodes of $G$. 
Exploiting all these properties, we observe that for $\kappa > 0$, 
$GF$ is positive at small $r$ and negative at large $r$, and hence has
an odd number of zeroes. The first and last zeroes 
of $GF$ correspond to nodes of $F$ and 
since the nodes of $F$ and $G$ alternate \cite{levgino2,rose1}, then 
the number of nodes of $F$ exceed by one the number of nodes 
of $G$. On the other hand, for $\kappa < 0$, $GF$ has the same 
(negative) sign near both end points, and hence has an even number of 
zeroes. By similar arguments we find that in this case the first and 
last zeroes of $GF$ are nodes of $G$ and $F$ respectively, and 
that $G$ and $F$ have the same number of nodes. These properties of 
the radial wave functions are illustrated in Figs.~1a-d for the 
$2p_{3/2}$ $(\kappa=-2)$, $1f_{5/2}$ $(\kappa=3)$ states. 
Altogether we have,
\ba
\begin{array}{ll}
\kappa < 0: \quad & n_F = n_G \qquad \\ 
\kappa > 0: \quad & n_F = n_G + 1 
\end{array}
\label{nodes}
\ea
where $n_F$ and $n_G$ denote the number of internal nodes of $F$ and $G$.  
\noindent
\begin{figure}
\begin{minipage}{0.48\linewidth}
\epsfig{file=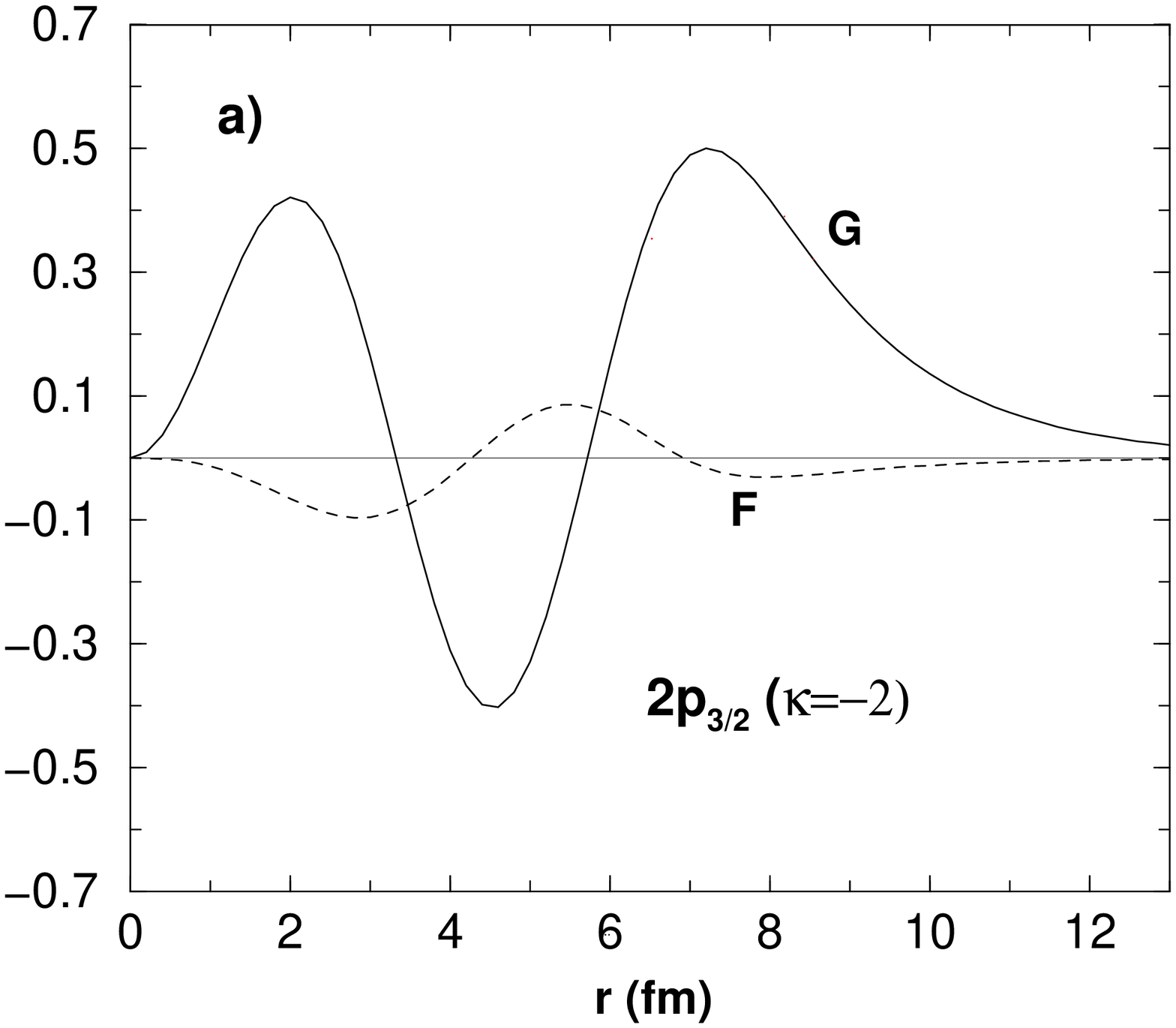,height=55mm,width=\linewidth,angle=-90}
\end{minipage}
\hspace{\fill}
\begin{minipage}{0.48\linewidth}
\epsfig{file=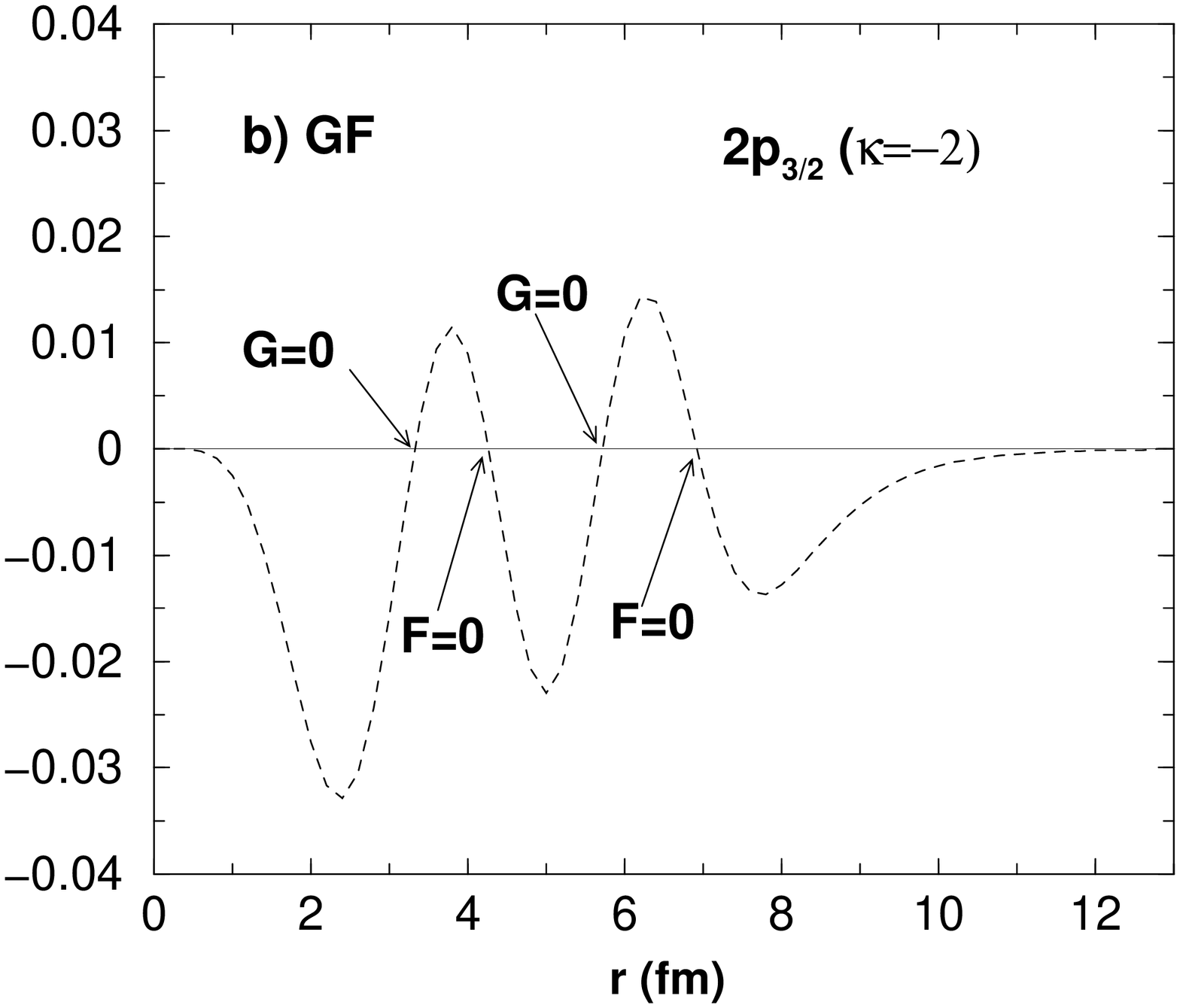,height=55mm,width=\linewidth,angle=-90}
\end{minipage}
\begin{minipage}{0.48\linewidth}
\epsfig{file=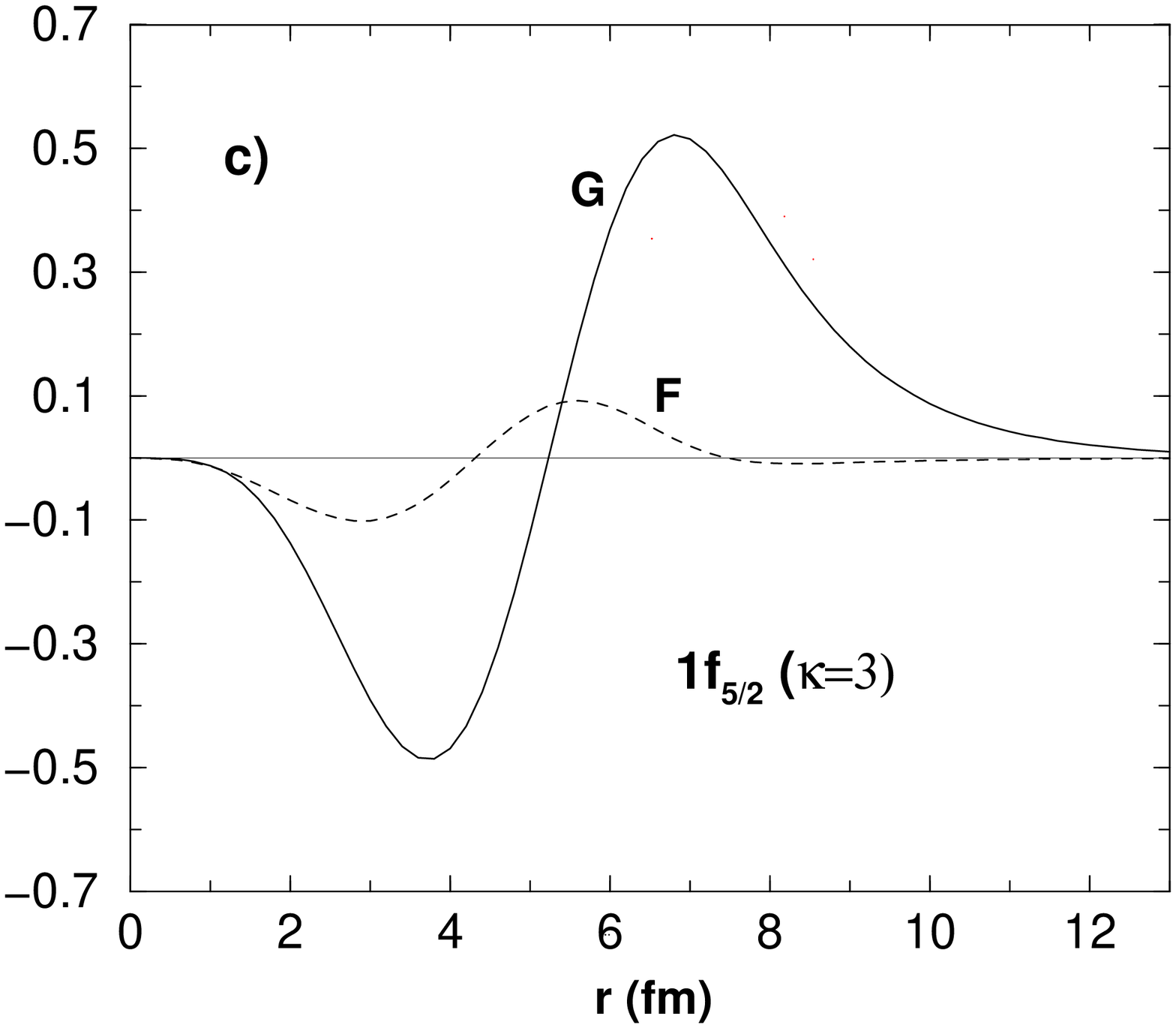,height=55mm,width=\linewidth,angle=-90}
\end{minipage}
\hspace{\fill}
\begin{minipage}{0.48\linewidth}
\epsfig{file=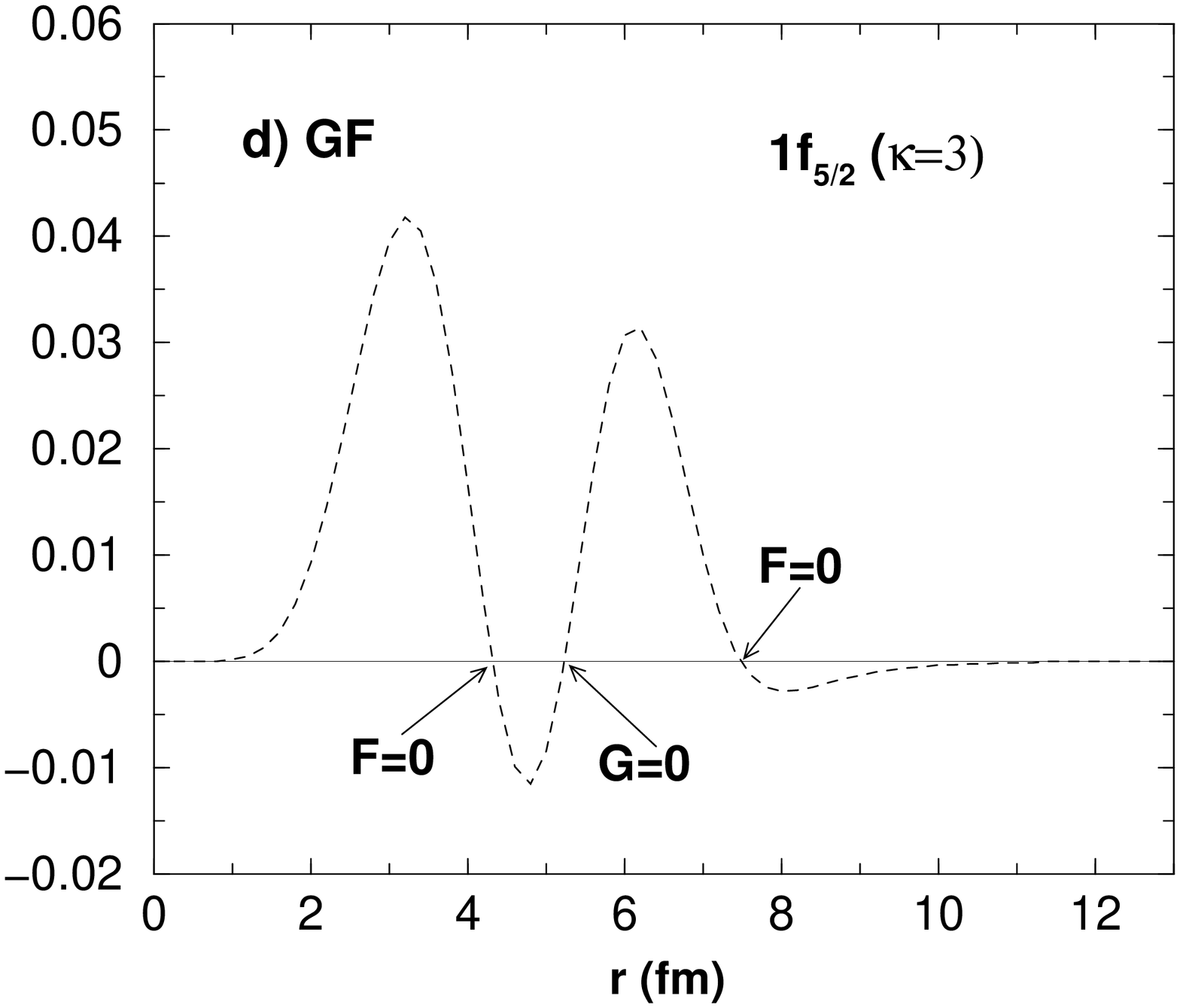,height=55mm,width=\linewidth,angle=-90}
\end{minipage}
\begin{minipage}{0.48\linewidth}
\epsfig{file=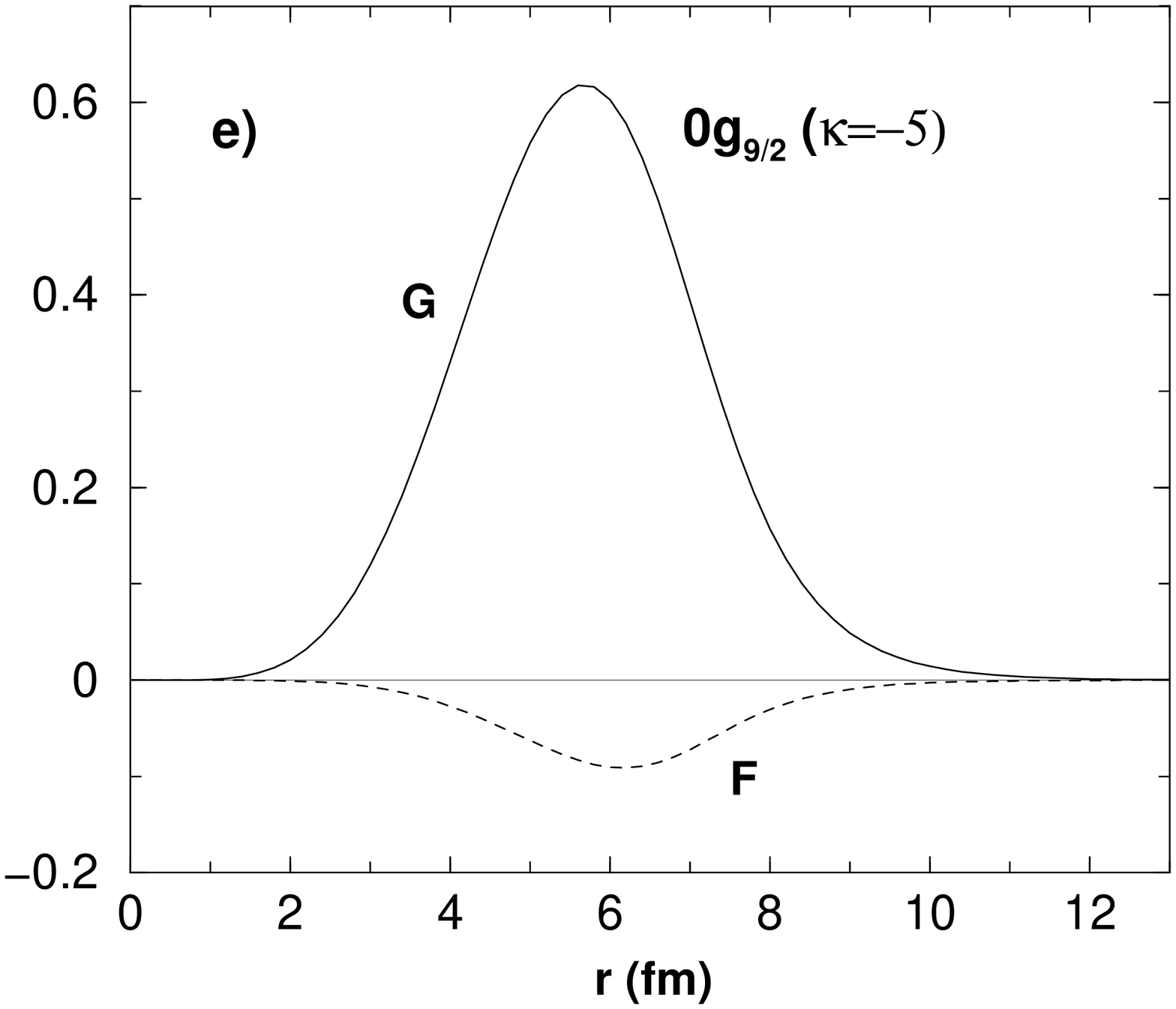,height=55mm,width=\linewidth,angle=-90}
\end{minipage}
\hspace{\fill}
\begin{minipage}{0.48\linewidth}
\epsfig{file=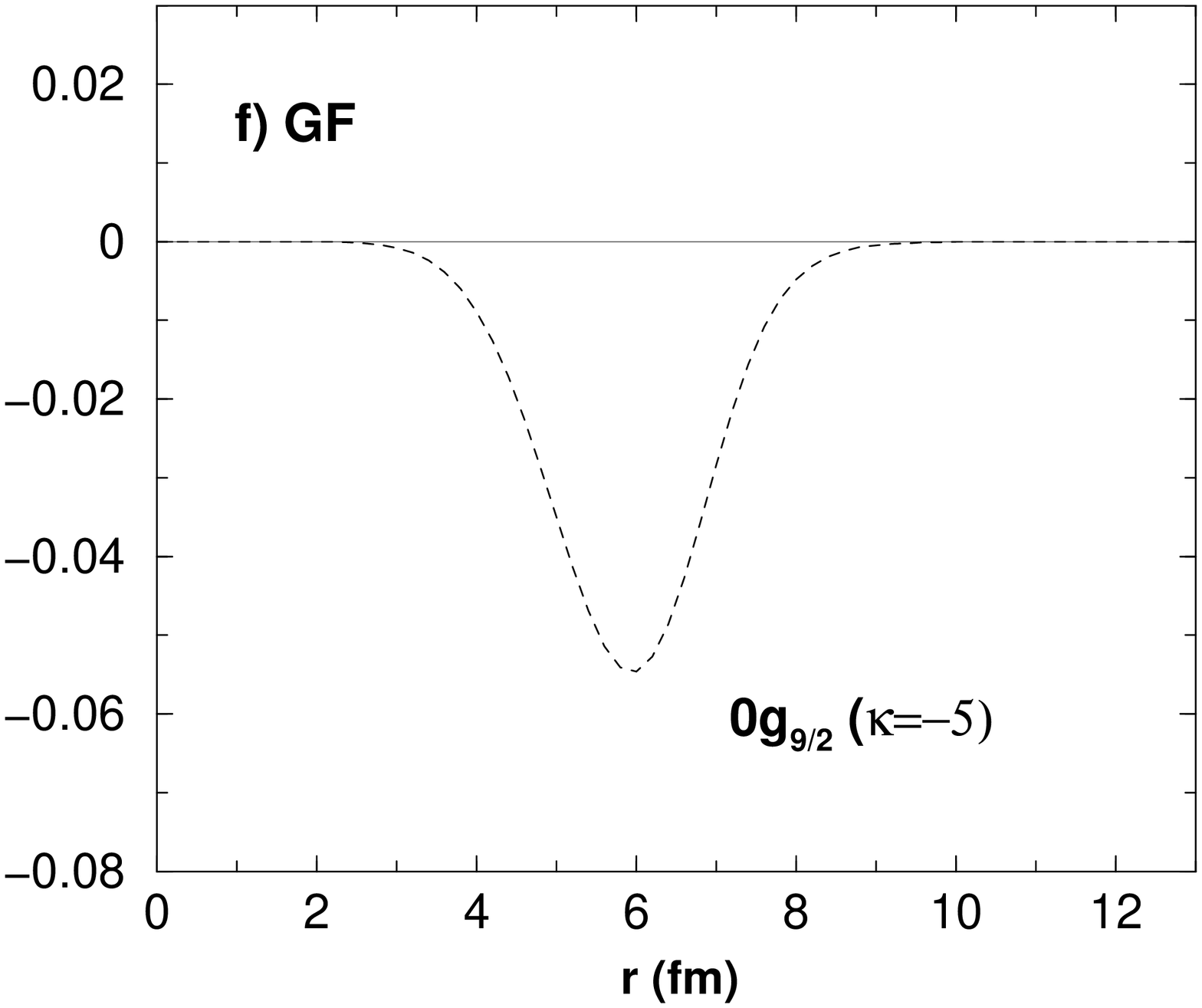,height=55mm,width=\linewidth,angle=-90}
\end{minipage}
\caption{(a) The radial upper component ($G$) and lower component 
($F$) in (Fermi)$^{-1/2}$ and (b) the corresponding 
product $GF$ in (Fermi)$^{-1}$ 
of the $2p_{3/2}$ $(\kappa=-2)$ state. 
(c) and (d) The same for the $1f_{5/2}$ $(\kappa=3)$ state. 
(e) and (f) The same for the $0g_{9/2}$ $(\kappa=-5)$ state.
All states are eigenstates of a Dirac Hamiltonian with scalar ($S$) and 
vector ($V$) potentials: $V_{S,V}(r)= 
\alpha_{S,V}\left [1+ \exp({r-R\over a})\right]^{-1}$ 
with parameters: $\alpha_S = -358$, $\alpha_{V}= 292$ MeV, $R=7$ fm, 
$a=0.6$ fm, tuned to the neutron spectra of $^{208}$Pb. 
Taken from Ref. \protect\cite{levgino2}.}
\end{figure}

When the radial parts of the wave 
functions $G$ and $F$ have no nodes ($n_F=n_G=0$), the corresponding 
bound states can appear only in the $j=\ell+1/2$ state ($\kappa < 0$), 
but not in the $j=\ell-1/2$ state ($\kappa > 0$) \cite{levgino2,hirooka}. 
The reason being that in this case  $GF$ has the same sign for all $r$ 
and since always $GF<0$ at large $r$, then it is also negative 
at small $r$ and therefore must have $\kappa < 0$. 
An example is shown in Figs.~1e-f for the case of the 
$0g_{9/2}$ $(\kappa=-5)$ intruder state.

Finally, we note that for bound states to exist,  
there must be a region where $B(r)>0$. This is due to 
the fact that, as mentioned, $GF$ is an increasing negative function at 
large $r$, and to enable it to vanish at $r=0$, its derivative 
$\left ( GF \right )'$ must change sign from positive to negative in 
some region. A glance at Eq.~(\ref{GFprime}) shows that since $A(r)>0$, 
a necessary (but not sufficient) condition for $\left ( GF \right )' $ 
to become negative is that 
\ba
B(r) = \left [\,E - M - V_S(r) - V_V(r) \, \right ] > 0\;\; 
{\rm for\; some\; r}~.
\label{delta}
\ea
The above condition means 
that in order that bound states exist, there has to be a region where the 
depth of the average attractive single-nucleon potential, $V_S(r)+V_V(r)$, 
is larger than the binding energy.  

\section{Relativistic Pseudospin Symmetry}

All of the above results are relevant for understanding properties of 
states in the relativistic pseudospin scheme \cite{gino,ami}. 
When $V_S= -V_V$, the Dirac Hamiltonian is a scalar under an SU(2) 
algebra, $[\,H\,,\, \hat{\tilde{S}}_{\mu}\,] = 0$. 
If in addition the potentials are spherically symmetric, 
the Dirac Hamiltonian has an additional invariant
SU(2) algebra, $[\,H\,,\, \hat{\tilde{L}}_{\mu}\,] = 0$. 
The relativistic pseudospin generators, $\hat{\tilde{S}}_{\mu}$, 
and relativistic pseudo-orbital angular momentum operators, 
$\hat{\tilde{L}}_{\mu}$, are given by
\ba
{\hat{\tilde {S}}}_{\mu} =
\left (
\begin{array}{cc}
\hat {\tilde s}_{\mu} &  0 \\
0 & {\hat s}_{\mu}
\end{array}
\right ) \;\; , \;\;
\hat{\tilde{L}}_{\mu} =
\left (
\begin{array}{cc}
\hat {\tilde \ell}_{\mu} &  0 \\
0 & {\hat \ell}_{\mu}
\end{array}
\right ) ~.
\label{SLgen}
\ea
Here $\hat {\tilde s}_{\mu} = U_p\, {\hat s}_{\mu}$ $U_p$ and 
$\hat {\tilde \ell}_{\mu} = U_p\, {\hat \ell}_{\mu}$ $U_p$, where
${\hat s}_{\mu} = \sigma_{\mu}/2$ and 
${\hat\ell}_{\mu} = \mbox{ \boldmath $r$}\times \mbox{ \boldmath $p$}$ 
are the usual spin and orbital angular momentum operators respectively, 
$\sigma_{\mu}$ the Pauli matrices and 
$U_p = \, {\mbox{\boldmath $\sigma\cdot p$} \over p}$ is the
momentum-helicity unitary operator introduced in \cite {draayer}. 
The sets $\left\{\hat{\tilde S}_{\mu},\;
\hat{\tilde s}_{\mu},\;\hat{s}_{\mu}\right\}$ 
and $\left \{\hat{\tilde L}_{\mu},\; \hat{\tl}_{\mu},\; 
\hat{\ell}_{\mu}\right\}$ 
form two triads of $SU(2)$ algebras. 
The $\hat{\tilde S}_{\mu}$ and $\hat{\tilde L}_{\mu}$ operators 
act on the four-components Dirac wave functions. 
The ${\hat{\tilde s}}_{\mu}$ and ${\hat{\tilde l}}_{\mu}$ operators form 
the non-relativistic pseudospin and pseudo-orbital angular momentum 
algebras respectively, and act on the upper components of the 
Dirac wave functions. The ${\hat s}_{\mu}$ and ${\hat \ell}_{\mu}$ 
act on the ``small'' lower components of the Dirac wave functions. 
The pseudospin ${\tilde s}=1/2$ and pseudo-orbital angular momentum $\tl$ 
are seen from Eq.~(\ref{SLgen}) to be the ordinary spin and 
ordinary orbital angular momentum of the lower component of the Dirac wave 
functions. The two states in the pseudospin doublet share a common 
pseudo-orbital angular momentum $\tl$, which is coupled to a 
pseudospin $\tilde{s}=1/2$, and thus have the form 
\ba
\Psi_{\kappa<0,m} &=& 
\left ({G_{\kappa<0}(r)\over r}
[Y_{\tl-1}\,\chi]^{(j)}_m\,,\,
i{F_{\kappa<0}(r)\over r}
[Y_{\tl}\,\chi]^{(j)}_m \right) \quad\;\; 
j =\tl-{1\over 2}\quad 
\nonumber\\
\Psi_{\kappa^{\prime}>0,m} &=& 
\left ({G_{\kappa^{\prime}>0}(r)\over r}
[Y_{\tl+1}\,\chi]^{(j^{\prime})}_m\,,\,
i{F_{\kappa^{\prime}>0}(r)\over r}
[Y_{\tl}\,\chi]^{(j^{\prime})}_m \right) \; 
j^{\prime}=\tl+{1\over 2} ~.\quad\;
\label{reldoub}
\ea
The underlying Dirac structure ensures that the wave function of the 
upper component of the Dirac eigenfunction has a spherical harmonic of 
rank $\ell= \tl-1$ for aligned spin: $j=\tl-1/2=\ell+1/2$, and a 
spherical harmonic of rank $\ell+2=\tl+1$ for unaligned spin: 
$j^{\prime}=\tl +1/2=(\ell+2) - 1/2$. This explains the particular angular 
momenta defining the pseudospin doublet of Eq.~(\ref{psdoub}). 
In the pseudospin symmetry limit the two states in Eq.~(\ref{reldoub}) 
form a degenerate doublet ($S=1/2$), and are connected by the pseudospin 
generators $\hat{\tilde S}_{\mu}$ of Eq.~(\ref{SLgen}). 
The corresponding upper components are a doublet 
with respect to the set $\hat{\tilde s}_{\mu}$ 
(the non-relativistic pseudospin algebra). Since the latter, 
by definition, intertwine space and spin, they can connect states for 
which the upper components have different radial wave functions, 
$G_{\kappa<0}(r)\neq G_{\kappa^{\prime}>0}(r)$. On the other hand, 
the corresponding lower components are a doublet with respect 
to the ordinary spin ${\hat s}_{\mu}$, and hence, in the pseudospin limit, 
their radial wave functions are equal up to a phase, 
\ba
F_{\kappa<0}(r) = F_{\kappa^{\prime}>0}(r) ~.
\label{twof}
\ea
In particular, $F_{\kappa<0}(r)$ and $F_{\kappa^{\prime}>0}(r)$ 
have the same number of nodes, which we denote by $n_r$. If we now use 
the result of Eq.~(\ref{nodes}), we find for $n_r\neq 0$ that 
$G_{\kappa<0}(r)$ in Eq.~(\ref{reldoub}) has also $n_r$ radial nodes, 
while $G_{\kappa^{\prime}>0}(r)$ has $n_r-1$ nodes. 
This explains the structure of nodes in the 
pseudospin doublet of Eq.~(\ref{psdoub}). 
The simple relation in Eq.~(\ref{twof}) between the radial wave functions 
of the lower components of the two states in the doublet, dictates 
this particular relation between the radial nodes of the corresponding  
upper components. This result cannot be obtained if one considers just 
the non-relativistic pseudospin algebra, $\hat{\tilde s}_{\mu}$, and is a 
direct outcome of the behavior of nodes of Dirac bound states 
and the identification of pseudospin as a relativistic symmetry 
of the Dirac Hamiltonian.

As we have shown, bound Dirac states, for which both the upper and 
lower components have no nodes ($n_r=0$) can occur only for 
$\kappa < 0$ and not for $\kappa^{\prime}> 0$. From 
Eq.~(\ref{reldoub}) we find that such states have pseudo-orbital 
angular momentum $\tl$ and total angular momentum $j = \tl -1/2$. 
As mentioned, these intruder states are ignored in the non-relativistic 
pseudospin scheme, and it is only the relativistic interpretation of
pseudospin symmetry, combined with known properties of Dirac bound 
states, which enable a classification for these states, as well as 
provide a natural explanation why these states do not have a 
pseudospin partner which is an eigenstate of the Hamiltonian.

The exact pseudospin limit requires that $V_S(r) = -V_V(r)$, which 
implies that $B(r) = E-M$. It is clear that under such 
circumstances the condition of Eq.~(\ref{delta}) cannot be fulfilled 
for bound states with positive binding energy $M-E>0$. This explains 
why in the exact pseudospin limit, there are no bound Dirac states 
and, therefore, by necessity the pseudospin symmetry must be broken 
in nuclei. Nevertheless, a variety of realistic mean field calculations
show that the required breaking of pseudospin symmetry in nuclei is 
small \cite{gino2,ring,arima,ginoami}. 
Quasi-degenerate doublets of normal-parity states, and abnormal-parity 
levels without a partner eigenstate persist in the spectra, and the 
relation of Eq.~(\ref{twof}) is obeyed to a good approximation, 
especially for doublets near the Fermi surface. 
As discussed, these features are sufficient to ensure the observed 
structure of nodes occurring in pseudospin doublets 
and the special status of intruder levels in nuclei.

\section{Acknowledgements}

It is a pleasure to dedicate this article to P. Ring on the occasion of 
his 60th birthday.
This research was supported in part by the U.S.-Israel Binational Science 
Foundation and in part by the United States Department of Energy under 
contract W-7405-ENG-36.

\end{document}